%% file: main.tex
\documentclass{aastex631}

\newcommand{\rv}{$R_{\rm V}$}
\newcommand{\av}{$A_{\rm V}$}

\newcommand{\pebv}{$P_{\rm V}/E(B-V)$}
\newcommand{\pv}{$P_{\rm V}$}

\begin{document}

\title{Extreme Starlight Polarization Efficiency Toward $\zeta$ Ophiuchi: A Case for Line-of-Sight Foreground Subtraction}

\correspondingauthor{Jordan A. Bartlett}
\email{jbartle9@uwyo.edu}

\author[0000-0002-9285-5369]{Jordan A. Bartlett}
\affiliation{Department of Physics \& Astronomy, University of Wyoming, 1000 E. University Ave., Laramie WY 82071, USA}

\author[0000-0002-4475-4176]{Henry A. Kobulnicky}
\affiliation{Department of Physics \& Astronomy, University of Wyoming, 1000 E. University Ave., Laramie WY 82071, USA}



\begin{abstract}

Despite the pervasive nature of interstellar dust and its impact on nearly all observations, most dust corrections employ line-of-sight averages over large angular scales. This neglects real variations on small angular and distance scales from discrete components of the ISM. We use $V$ band polarimetry, public dust maps, and \textit{Gaia} DR3 distances of 25 stars along a 50\arcmin\ radius sight line towards the O9.5IV star $\zeta$ Ophiuchi ($d \approx $ 182 pc) to examine both dust and magnetic structures over the range $d = $ 36--1176 pc and angular scales of $< 1$\degr. Polarization and reddening data indicate two discrete dust populations having different magnetic field orientations along the sight line, one at $d \simeq$ 86--127 pc and another at $d\simeq$ 252--287 pc. After removal of the foreground, the more distant component exhibits alignment in polarization angle with 12 \micron\ PAH striations seen in the field. This more distant dust population exhibits evidence of extreme starlight polarization efficiency with an average of 14.1\% mag$^{-1}$, greater than the canonical Serkowski limit of 9\% mag$^{-1}$. The spatial coincidence with the PAH striations indicates the PAH-emitting grains and those responsible for the high polarization efficiency may be components of the same dust population. We find no evidence that $\zeta$ Oph's radiative influence affects the polarizing or reddening properties of the surrounding dust. Our study demonstrates that accurate distance-based foreground subtraction is vital to properly understanding superimposed dust and the magnetic field components in the ISM.

\end{abstract}



\section{Introduction} \label{sec:intro}

Dust grains, primarily carbonaceous and siliceous materials on the order of 0.01--0.4 \micron\ in diameter, are prevalent throughout the interstellar medium (ISM). These grains leave their imprint on starlight by simultaneously reddening, extincting, and polarizing the light \citep[reviews by][]{savagemathis1979, mathis1990, draine2003}. The relative magnitude of each phenomenon encodes information about the size, composition, orientation, and alignment of interstellar grains.

The ratio of extinction to reddening, $R_{\rm V} = A_{\rm V} / E(B-V)$ is commonly used to characterize the relative impact of these two phenomena. The Milky Way average is \rv\ = 3.1, but there is significant evidence to suggest that \rv\ varies from $\approx$2--6 depending on the sight line \citep{johnsonmorgan1955, ccm89, draine2003}. In dense regions of the ISM, where larger grains can grow without being destroyed, \rv\ is typically larger ($> 3.1$), as these grains produce more extinction per unit of reddening. Accretion and coagulation of grains, not just a lack of grain destruction, within these dense areas also contribute to a higher relative population of large grains \citep{hirashita2012}. Smaller \rv\ ($< 3.1$) indicates a greater percentage of smaller dust grains which elevates reddening relative to extinction. Highly irradiated areas of the diffuse ISM are more likely to contain these populations with inflated fractions of small dust grains, as they are more exposed to sputtering radiation \citep{mathis1977, aannestad1995, weingartnerdraine2001}. Grain-grain shattering can also contribute to a shift in the dust size distribution towards smaller grains, preserving overall dust mass but increasing the number of small grains that lead to more reddening per unit extinction \citep{mckinnon2018}.

Reddening curves have been developed to quantify the amount of reddening or extinction as a function wavelength. One way to characterize these curves is through a ratio of color excesses: $k(\lambda-V) = E(\lambda-V)/E(B-V)$ \citep{bless1970, massa1986, fitzpatrick1988, mathis1990}. There are a variety of successful reddening curve parameterizations that express the variation in dust content exhibited by Galactic sight lines, but the most popular over the past few decades are parameterized by the single variable \rv\ \citep[e.g.,][hereafter, G23]{ccm89, f99, f19, g23}. The availability of all-sky surveys that provide photometric measurements and stellar parameters for hundreds of millions of stars make it possible to map dust extinction \citep{green2019, edenhofer2024, zucker2025} and measure reddening curves \citep{schlafly2016, zhang2025} over arbitrarily large or small angular scales and as a function of distance in the Milky Way.

Since \citet{hiltner1949} and \citet{hall1949} discovered that dust grains in the ISM cause polarization of background starlight, this effect has been used to trace the structure of ambient magnetic fields. \citet{davisgreenstein1951} first proposed the process of paramagnetic relaxation as a method of grain alignment, in which non-spherical, rotating dust grains within a magnetic field align with their long axes perpendicular to the direction of the \textbf{B} field. Growing evidence indicates that Radiative Torques (RAT) are additionally necessarily to align dust grains to the degree that is required by observed polarization \citep{dolginovmitrofanov1976, draineweingartner1996, draineweingartner1997, lazarianhoang2007}. The resulting polarization of the background starlight maps the projected orientation of the ambient \textbf{B} field, under the assumption that no other environmental factors influence grain alignment. \citet{Serkowski1975} introduced the concept of polarization efficiency, the ratio of $V$-band polarization percentage (\pv) to $E(B-V)$. They also proposed that there is an upper limit to polarization efficiency, with their empirical upper bound being \pebv\ $\leq$ 9\% mag$^{-1}$. More recently, studies have suggested that the upper limit could be at least \pebv\ $\leq $ 13--18\% mag$^{-1}$ \citep{planck2015, panopoulou2019}. The physical origins of these high-efficiency polarizing regions has not been conclusively identified but may be linked to a fortuitous projection of the magnetic field in the plane of the sky \citep{planck2015, panopoulou2019}.

One complication in mapping ambient \textbf{B} fields through polarimetry, as well as determining the polarization efficiency, arises from the possible superposition of multiple dust and magnetic structures along a single line of sight. There is evidence to suggest that the structures of ambient \textbf{B} fields vary on distance scales of tens to hundreds of parsecs \citep{hildebrand2009, jones1989}. Overlapping magnetic structures along a single sight line can either more highly polarize or depolarize starlight depending on whether the components are aligned or anti-aligned. In contrast, multiple dust components can only increase the observed reddening or extinction. \cite{skalidis2024} suggests that many regions within the galaxy may have super-Serkowski polarization efficiencies (i.e., efficiencies greater than 9\% mag$^{-1}$), but line-of-sight superpositions of multiple magnetic components may lead to depolarization and lower observed efficiencies. In order to recover the ambient \textbf{B} field or polarization efficiency along any sight line, foreground components must first be removed. This requires target stars with a range of precisely known distances, something only recently possible with high quality parallax data \citep{GaiaMission}.

\citet[][hereafter, PK22]{picconekobulnicky2022} investigated the line of sight centered on the O9.5IV star $\zeta$ Ophiuchi ($\zeta$ Oph), at a Heliocentric distance of $d = 182$ pc \citep{Bailer-Jones2018}, in order to test whether a highly irradiating source could produce a measurable impact on the dust grains in its immediate vicinity. There is a rich body of literature focusing on characterizing the diffuse dust clouds in the foreground of $\zeta$ Oph \citep[ex.][]{Serkowski1975, draine1986, ccm89, tachihara2000, liszt2009, choi2015, siebenmorgen2020}, making it a promising sight line to understand the effects of environment on dust. PK22 measured 27 background stars ($d = $ 203--8434 pc,  $d_{\rm median} = $ 582 pc) using optical $V$ band polarimetry as well as spectrophotometry. Through progressive foreground subtraction, they found evidence to suggest the presence of at least three separate dust clouds along the line of sight: foreground dust ($< 200$ pc), highly-polarizing mid-distance dust (200--300 pc), and non-polarizing distant dust (600--2000 pc). PK22 also posited that the low-\rv, highly polarized dust they found within the mid-distance cloud could be a result of the dust grains being sputtered and aligned by the radiation from $\zeta$ Oph. When foreground-subtracted with the polarization of $\zeta$ Oph, the projected \textbf{B} field permeating this mid-distance dust also aligns with PAH striations present in the 8 \micron\ Spitzer \citep{werner2004} image of the field. They also found that many sight lines within this cloud fall above the traditional Serkowski \pebv\ = 9\% mag$^{-1}$. However, the PK22 sample was limited to target stars substantially in the background of $\zeta$ Oph ($d > 200$ pc), making it difficult to determine the location of this highly-polarizing dust, as well as limiting their ability to link the characteristics of the dust to the radiative influence of $\zeta$ Oph. Furthermore, their estimations of $E(B-V)$ were limited by the availability of fundamental stellar parameters for their target stars.

In this work, we present new polarimetric data along 25 sight lines to stars in the immediate foreground and background of $\zeta$ Oph ($d = $ 36--1176 pc, median distance of 309 pc) with the goal of characterizing the PK22 foreground and mid-distance dust clouds on finer scales. We utilize similar line-of-sight depolarization and dereddening techniques to PK22 to identify the location of the highly-polarizing dust. In Section \ref{sec: targets}, we introduce the target sample and the naming conventions for the target groups used throughout the text. Section \ref{sec: pols} presents our new polarimetric data and how it varies with Heliocentric and angular distance. Section \ref{sec: red} describes our utilization of the \citet{zhang2025} dust maps for determining $E(B-V)$ and \rv. Section \ref{sec: discussion} provides an analysis of the physical implications of these data, how our results differ from PK22, and additional evidence supporting high polarization efficiency in the 250--400 pc distance range. Section \ref{sec: conclusions} summarizes the conclusions of our study and presents ideas for paths forward.

\section{Target Selection} \label{sec: targets}

We selected 25 stars brighter than approximately $V$ = 14 mag. within a 50\arcmin\ radius of $\zeta$ Ophiuchi. We used \textit{Gaia} DR3 \citep{GaiaMission,GaiaDR3} parallaxes to choose stars with a range of distances between 36 pc and 1176 pc ($d_{\rm median} =$ 309 pc) along the line of sight. This allowed us to probe the interstellar medium in the foreground and immediate background to $\zeta$ Oph \citep[$d = 182^{+53}_{-34}$ pc,][]{Bailer-Jones2018}. Table \ref{tbl: gaia} lists the targets and fundamental stellar parameters. The numbering system for the target stars, as listed in Column 1, is a continuation of the target numbers used in PK22 which increase in number with R.A. The targets in this paper begin at number 28 and continue through 52. Column 2 is the distance group in which the target has been sorted, as described later in this section. Column 3 is the \textit{Gaia} identifying number. Columns 4 and 5 present the \textit{Gaia} J2016 right ascension and declination of the targets in degrees. Column 6 is the  \textit{Gaia} DR3 inverse parallax distance.\footnote{Given that the fractional distance uncertainties are less than 1\% in most cases, we adopt a mean symmetrical uncertainty instead of asymmetric uncertainties.} Columns 7, 8, and 9 are the stellar parameters temperature, surface gravity, and metallicity derived from the \textit{Gaia} DR3 General Stellar Parametrizer-spectroscopy module (GSP-Spec; reported as {\tt teff\_gspspec}, {\tt logg\_gspspec}, and {\tt mh\_gspspec}). We adopt these rather than the photometrically determined parameters (GSP-Phot) because of the documented degeneracy\footnote{\url{https://gea.esac.esa.int/archive/documentation/GDR3/Data_analysis/chap_cu8par/sec_cu8par_apsis/ssec_cu8par_apsis_gspphot.html}} between effective temperature and extinction in the latter. Any targets missing GSP-Spec parameters are supplemented with GSP-Phot, as indicated by table footnotes. Finally,  Column 10 is the \textit{Gaia} DR3 Renormalized Unit Weight Error (RUWE) value.

\begin{table}[h]
    \centering
    \caption{Targets Observed and Stellar Parameters \label{tbl: gaia}}
    \input{table1}
    \tablenotetext{a}{No data from the GSP-Spec method were available. These data are instead taken from the GSP-Phot method.}
\end{table}

The majority of the target stars lie at distances between approximately 150 pc and 400 pc. We include targets from PK22 within this distance range (Targets 4, 6, 11, 18, and 24) in the analysis as additional data points as well as a source of comparison. The targets are sorted into three groups based on distance (as well as grouping in the Stokes $q$--$u$ plane, discussed in Section \ref{sec: pols}), designated as: the ``Near Group" at $d \leq 86$ pc (4 targets), the ``$\zeta$ Oph Group" at $127 \leq d \leq 252$ pc (8 targets), and the ``Far Group" at $d \geq 278$ pc (18 targets). These distance groups are not to be confused with the different distance groups of PK22, as they are used in a similar fashion to designate potentially discrete dust populations.

\section{Polarization} \label{sec: pols}

\subsection{Polarization Observations} \label{subsec: pol obs}
We conducted $V$-band polarimetric measurements of the 25 target stars with the OptiPol \citep{Jones2008} optical polarimeter on the Wyoming Infrared Observatory (WIRO) 2.3 m telescope on 2021 May 5 and 12. The OptiPol instrument contains a half-wave plate, which rotates between $22.5 \degr$, $45 \degr$, $67.5 \degr$, and $90 \degr$, as well as a Wollaston prism, which separates the incoming light into two orthogonal polarizations. OptiPol utilizes a $4096^2$ Apogee Alta F16M with Kodak KAF-16803 CCD with 9 $\micron$ pixels. The field of view is $40\arcsec$ x $80\arcsec$ with a plate scale of $0.\arcsec12$ pixel$^{-1}$ using 4 x 4 binning. We observed each star with a series of exposures, ranging between 5 s and 120 s, at each of the four half-wave plate positions. This results in a total of eight measured polarization angles for each target. For the initial processing of the images, we used standard IRAF \citep{tody1986} techniques to perform bias removal, dark current subtraction, and flat-fielding using exposures from a dome screen illuminated by quartz lamps. 

The unpolarized standard stars HD 94851 \citep{Turnshek1990} and BD+33 2642 \citep{Schmidt1992}, as well as the polarized standards HD 155197 and VI Cyg 12 \citep{Schmidt1992}, were used to determine the instrumental normalized Stokes parameters and the position angle offset: $q_{inst}$ = -0.10, $u_{inst}$ = 0.28, and $P.A._{inst}$ = $-48.0$\degr. After subtracting $q_{inst}$ and $u_{inst}$ in the $q$--$u$ plane, we calculated the $q$ and $u$ values of our target stars using the technique described in \citet{Tinbergen2005}. We  used the normalized stokes parameters to determine the raw polarization percentage in the $V$-band ($P_{\rm V}'$) using the equations $P_{\rm V}'= \sqrt{q^2 + u^2}$ and $P.A.=\frac{1}{2} \arctan(\frac{u}{q})$. We applied standard error propagation methods to determine the uncertainties in the four polarization values: $q$, $u$, $P_{\rm V}'$, and $P.A$. Photon statistics represent the minimum level of uncertainties in $P_{\rm V}'$ and $P.A.$; therefore, we multiplied all of the propagated uncertainties by a uniform factor of 3.5 to account for the dominant source of noise---time-variable sky conditions. We determined this factor empirically from the ratio of the RMS in our three or more individual measurements of each target to the uncertainty derived from photon statistics. The median $1 \sigma$ uncertainties are $\sigma_q = \sigma_u = 0.05$\%. Because $P_{\rm V}'$ is the quadrature sum of $q$ and $u$, we also correct for the positive bias using
 \begin{equation} \label{equ: P bias}
    P_{\rm V} = \sqrt{P_{\rm V}'^2 - \sigma_P^2},
\end{equation}
where $\sigma_P$ is the uncertainty on $P_{\rm V}'$. 

Table \ref{tbl: pols} lists these polarization measurements for each target. Column 1 is the target number used in this paper. Columns 2 and 3 are the measured normalized Stokes parameters $q$ and $u$. Column 4 is the positive bias corrected polarization percentage, as described by Equation \ref{equ: P bias}. Column 5 is the measured position angle on the sky of the polarization. Columns 6 and 7 are the same as 4 and 5 but foreground-corrected for the average $q$ and $u$ of the $\zeta$ Oph Group (described further in Section \ref{subsec: fore-sub}). The targets are arranged in order of increasing distance, rather than target number as in Table \ref{tbl: gaia}, and the horizontal lines delineate the three distance groups: Near (top), $\zeta$ Oph (middle), and Far (bottom).

\begin{table}[h]
    \caption{Raw and Foreground-Subtracted Polarizations \label{tbl: pols}}
    \begin{center}
    \input{table2}
    \tablenotetext{}{Horizontal lines delineate the three distance groups: Near, $\zeta$ Oph, and Far.}
    \end{center}
\end{table}

\begin{figure}[h]
            \plotone{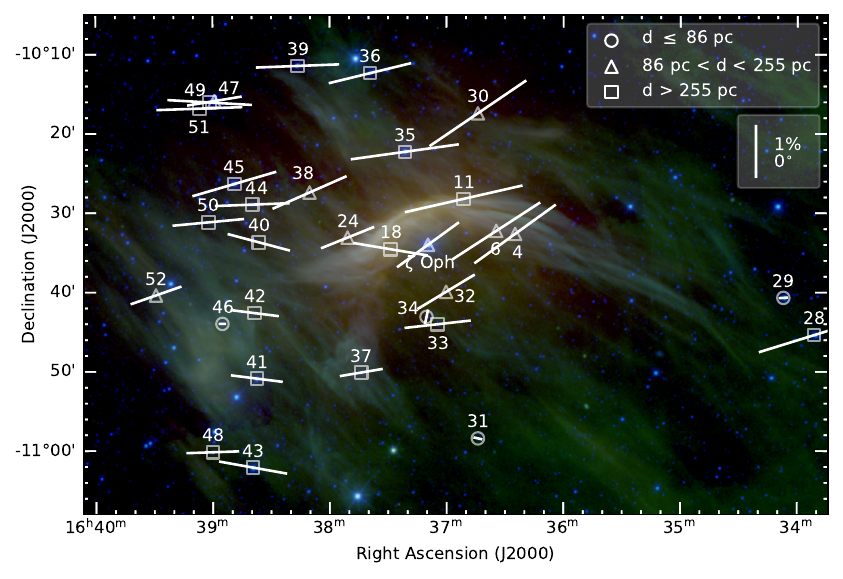}
            \caption{An RGB image of the target field in equatorial coordinates at 22 $\micron$, 12 $\micron$, and 3.4 $\micron$ represented as red, green, and blue, respectively. The lengths of the polarization line segments indicate the values of \pv, with 1\% polarization equal to the length indicated in the legend. Similarly, the angles of the line segments indicate the $P.A.$ of the targets, where $P.A. = 0$\degr\ is North. The symbols indicate the three distance groups of target stars, as depicted in the legend.}
    \label{fig: skypol}
\end{figure}

Figure \ref{fig: skypol} shows the polarization data for each target represented on the sky, in which the length and orientation of the line segments represent \pv\ and $P.A.$, respectively. The background image is an RGB combination of  22 $\micron$, 12 $\micron$, and 3.4 $\micron$ images from the \textit{Wide-field Infrared Survey Explorer} \citep[\textit{WISE;}][]{wright2010}. The well-known bowshock preceding the runaway star $\zeta$ Oph is prominent in the center of the image. It also shows the green 12 $\micron$ filamentary striations, linked to polycyclic aromatic hydrocarbon (PAH) emission, also seen at 8 \micron\ as noted by PK22, oriented at $P.A. \approx 54$\degr. The circle, triangle, and square symbols represent the Near, $\zeta$ Oph, and Far distance groups, respectively. We include five stars from PK22, Targets 4, 6, 11, 18, and 24, because they fall within the distance range of the $\zeta$ Oph Group and increase the number of targets. We also include the polarization of $\zeta$ Oph, as presented in \citet{Serkowski1975}. \pv\ values range between $0.14\pm0.03$\% and $2.20\pm0.06$\%. The four stars in the Near Group are distinct, showing minimal polarization with \pv\ $< 0.28\pm0.10$\%. This indicates the Near Group is negligibly affected by aligned dust grains. Figure \ref{fig: skypol} shows that the $\zeta$ Oph and Far Groups have substantially greater polarization with \pv\ $> 0.82\pm0.08\%$. The targets in the $\zeta$ Oph Group have similar polarizations to $\zeta$ Oph itself which has \pv\ $\simeq 1.44\pm0.04$ and $P.A. \simeq 126\pm0.8$. The Far Group has similar \pv\ to the $\zeta$ Oph Group, but the $P.A.$s show a larger dispersion and a systematically smaller $P.A.$ by approximately 30\degr.

\subsection{The $q$--$u$ Plane} \label{subsec: qu}

\begin{figure}[h]
            \plotone{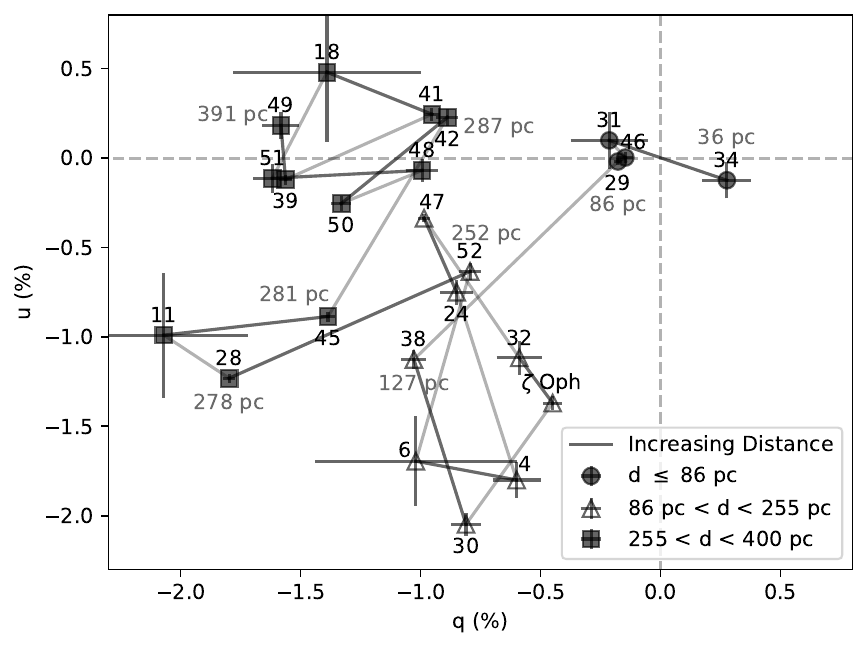}
            \caption{The target stars plotted in the Stokes $q$--$u$ plane. The shapes of the markers represent the three distance groupings: Near Group as circles, $\zeta$ Oph Group as triangles, and Far Group as squares. The gray line segments (alternating light and dark for visibility) connect the targets in order of increasing distance, beginning with Target 34 and ending with Target 49. For clarity, Target 24 has been offset by -1$\sigma$ in $q$, and Target 6 has been offset by -0.5$\sigma$ in $q$ and +0.5$\sigma$ in $u$.}
    \label{fig: qu-plot}
\end{figure}

Figure \ref{fig: qu-plot} presents the normalized Stokes parameters from Table \ref{tbl: pols} plotted in the $q$--$u$ plane. The targets represented in the figure are limited to those at distances less than 400 pc for clarity. The shape of the markers correspond to the Near, $\zeta$ Oph, and Far distance groups, with the same symbols as those in Figure \ref{fig: skypol}.  The gray connecting lines order the targets by increasing distance, beginning with target 34 (36 pc). The lines alternate dark and light gray in order to better view the sequence of targets. The black error bars represent the 1$\sigma$ uncertainties. The Near Group is tightly clustered near $q = u = 0\%$, consistent with minimal or no polarization out to a distance of 86 pc. The four targets in this group have nearly identical polarizations despite spanning nearly the entire approximately 1\degr\ field of view, indicating that the foreground polarizing component is mostly uniform on these angular scales. As the distance increases to that of the $\zeta$ Oph Group, beginning with target 38 (127 pc), there is a marked increase in the magnitudes of both $q$ and $u$ to $\vert q \vert = 0.76\%$ and $\vert u \vert = 1.22$\%. This indicates a dramatic jump in the polarization ($\overline{P_{\rm V}} = 1.49$\%, $\overline{P.A.} = 117.2$\degr) somewhere between 86 pc (target 29) and 127 pc (target 38). Within the $\zeta$ Oph Group, $q$ and $u$ change quasi-randomly, with standard deviations of $RMS_q = 0.18\%$ and $RMS_u = 0.55\%$, indicating stochastic magnetic field orientations within the distance range of 127--252 pc. The first three stars in the Far Group (Targets 28, 11, and 45) occupy a different locus in the $q$--$u$ plane, characterized by an increase in $\overline{P_{\rm V}}$ by $\Delta \overline{P_{\rm V}} = 0.54\%$, as well as a rotation in $\overline{P.A.}$ by $\Delta \overline{P.A.} = -11.8$\degr. After this transition, the remaining Far Group targets (beginning with Target 42 at a distance of 287 pc and ending with Target 49 at 391 pc) cluster together and show another marked systematic change, characterized by $\overline{P_{\rm V}} = 1.31$\% and $\overline{P.A.} = 88.1$\degr. This group is also visually distinct from the $\zeta$ Oph Group in the $q$--$u$ plane. Like the $\zeta$ Oph Group, this group shows small, seemingly stochastic, variations among the targets' $q$ and $u$ values. This subset of the Far Group will represent much of the analyses in Sections \ref{sec: red} and \ref{sec: discussion}.

\subsection{Structure Function} \label{subsec: struc func}

An examination of polarization of adjacent sight lines reveals coherent structure on small angular and line-of-sight distance scales. Comparing Figure \ref{fig: qu-plot} with Figure \ref{fig: skypol}, we find that there are some pairs of targets that have similar Heliocentric distances (small $\Delta d$), small angular separations ($\Delta \theta$) , and similar polarizations within the $q$--$u$ plane. The first of these pairs consists of $\zeta$ Oph and Target 32. They are separated by only $\Delta d = 1.6$ pc,  $\Delta \theta = 6.3$\arcmin\ (corresponding to a projected transverse separation $s = 0.33$ pc), and a $q$--$u$ vector magnitude distance,

\begin{equation}
    ||\Delta \vec{P_{\rm V}}|| = \sqrt{(\Delta q)^2 + (\Delta u)^2},
\end{equation}

\noindent of $||\Delta \vec{P_{\rm V}}|| = 0.29$\%. Another closely spaced pair is Target 4 and Target 6, separated by  $\Delta d = 28.2$ pc,  $\Delta \theta = 2.44$\arcmin\ ($s = 0.16$ pc), and $||\Delta \vec{P_{\rm V}}|| = 0.21$\%. Similarly, we find there are pairs of targets with small $\Delta d$, but have much larger $\Delta \theta$ and are markedly different in the $q$--$u$ plane. One such pair is Targets 6 and 52. These are separated by only $\Delta d = 0.1$ pc, but they have $\Delta \theta = 45$\arcmin\ ($s = 3.3$ pc) and $||\Delta \vec{P_{\rm V}}|| = 1.19$\%, suggesting that there are appreciable variations in the geometry of the polarizing component on few$\times$10\arcmin\ ($\sim$pc) scales. Another of these pairs consists of targets 48 and 51 which have $\Delta d = 1$ pc, but $\Delta \theta = 45$\arcmin ($s = 4.2$ pc). In the $q$--$u$ plane, they are separated by $||\Delta \vec{P_{\rm V}}|| = 0.60$\%. Both of these widely separated pairs have similar $P.A.$ but substantially different \pv, indicating a constant projected magnetic field orientation on the sky but very different polarizing power. The similarities and differences in polarization for these pairs suggest that the dust grain alignment distribution is uniform across small angular distances of $\Delta \theta < 10$\arcmin, transverse distance scales of $s < 0.4$ pc, and line-of-sight distances of $\Delta d < 30$ pc. 

\begin{figure}[h]
            \plotone{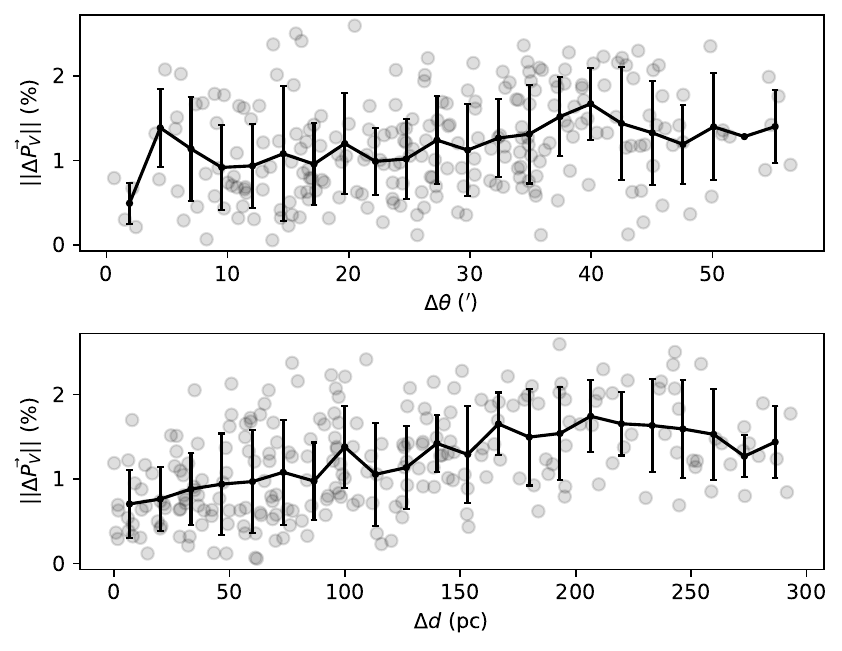}
            \caption{The structure function of $||\Delta \vec{P_{\rm V}}||$ for line of sight $\Delta d$ and angular separation $\Delta \theta$. The data are binned into 22 bins for clarity, with the black data points representing the mean value of each bin. The error bars represent the 1$\sigma$ uncertainties. These data are limited to $\Delta d < 300$ pc and $\Delta \theta < 1$\degr due to lack of target pairs beyond these separations. The un-binned data are presented as semi-transparent points.}
    \label{fig: structure func}
\end{figure}

In order to better quantify the angular and line-of-sight distances over which polarization properties are correlated, we introduced a modified version of the traditional structure/dispersion function \citep[see][]{falcetagoncalves2008,hildebrand2009,chapman2011}, in which we substitute $||\Delta \vec{P_{\rm V}}||$ for the change in $P.A.$ and bin the resulting values in $\Delta d$ and $\Delta \theta$. Because our structure function allows us to examine vector differences of points within the $q$--$u$ plane, it takes into account both changes in $P.A.$ and \pv\ as a function of angular separation and Heliocentric distance. Figure \ref{fig: structure func} displays the structure function for our targets with respect to $\Delta \theta$ (\textit{bottom panel}) and $\Delta d$ (\textit{top panel}), with 22 data bins encompassing 276 unique target pairs. The error bars represent the standard deviations within each bin. The presented data are limited to $\Delta d < 300$ pc and $\Delta \theta < 1$\degr\ because there are too few target pairs beyond these limits. We find that $||\Delta \vec{P_{\rm V}}||$ vs. $\Delta d$ has a quasi-linear increase (slope $\approx$ 0.004\% pc$^{-1}$) between $0\ \text{pc} < \Delta d < 200\ \text{pc}$, beyond which the data appear to level. For $||\Delta \vec{P_{\rm V}}||$ vs. $\Delta \theta$, we see an initial sharp increase between the first and second bins ($\Delta \theta \gtrsim 0.1$\degr), then there is a slow rise (slope $\approx 0.05$\% arcmin$^{-1}$) out to $\Delta \theta \simeq 40$\arcmin, beyond which the data appear constant. 

From these behaviors, we are able to draw conclusions about the scale lengths of these observed magnetic structures through the modified structure function. The scale on which the structure function becomes flat compared to $\Delta d$ indicates that the polarizing dust has a component that is mostly constant on larger scales of $\Delta d \gtrsim $ 150--200 pc. We also determine that the constant component of the polarization has a scale of $\Delta \theta \gtrsim 40$\arcmin, based on the location at which the data have a sharp increase before remaining mostly constant. On scales smaller than these, \pv\ and $P.A.$ may change in quasi-random ways, as we observed among distance groups in Figure \ref{fig: qu-plot}.

\subsection{Foreground Subtraction} \label{subsec: fore-sub}

In order to examine the effects of $\zeta$ Oph on the polarization properties of the dust in its vicinity, we need to isolate the polarization signature of that dust from the signature of the dust in the foreground. Therefore, we subtract off a foreground polarization from all of our data. We designate the foreground as the average $q$ and $u$ values of the $\zeta$ Oph Group, with $\bar{q} = -0.76\%$ and $\bar{u} = -1.22\%$. Because they are all within 100 pc of $\zeta$ Oph, and are the nearest targets that show significant polarization, we treat these as the population representing the first instance of polarizing dust along the sight line. Subtracting their average polarization allows us to better see the true polarization of the Far Group targets. We performed this foreground correction by subtracting $\bar{q}$ and $\bar{u}$ of the $\zeta$ Oph Group from all targets in the $\zeta$ Oph and Far Groups in the $q$--$u$ plane. This shifts the targets in the $q$--$u$ plane so the $\zeta$ Oph Group is clustered near $q = u = 0$ and the Far Group lies within the second quadrant ($P.A.$ between $45 \degr$--$90 \degr$). 

\begin{figure}[h]
            \plotone{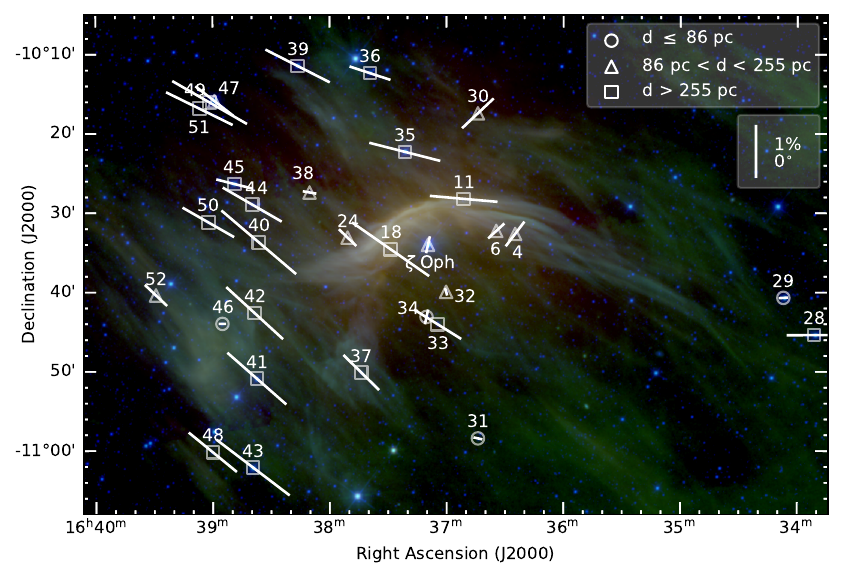}
            \caption{An RGB image of the foreground-corrected target field, formatted as in Figure \ref{fig: skypol}. The $\zeta$ Oph and Far Groups have been corrected by the average value of the $\zeta$ Oph Group ($\bar{q} = -0.76\%$ and $\bar{u} = -1.22\%$). Because they are nearer than the foreground, polarizing dust, the Near Group targets have not been corrected.}
    \label{fig: skypol-subfore}
\end{figure}

Figure \ref{fig: skypol-subfore} presents this foreground-subtracted polarization data on the sky, with the same formatting as in Figure \ref{fig: skypol}. The Near Group has not been foreground corrected, as they are nearer than the designated foreground dust. The values of \pv\ within the corrected $\zeta$ Oph Group are now much smaller, and they are more randomly oriented in $P.A.$ as the average in the $q$--$u$ plane is shifted to zero. The most notable part of Figure \ref{fig: skypol-subfore} is the Far Group, which now exhibits an average position angle of $\overline{P.A.} = 56.2$\degr. This average orientation of the polarization line segments now aligns with the previously noted 12 \micron\ PAH striations, which have an average orientation of $\overline{\theta} = 54.4$\degr (as calculated by PK22). This indicates that these striations lie beyond the $\zeta$ Oph Group, and are likely responsible for a significant portion of the polarizing dust for targets beyond $d \approx 255$ pc.

\section{Reddening Measurements} \label{sec: red}

We used the all-sky reddening catalog of \citet{zhang2025}\footnote{Catalogs and sample code available at https://doi.org/10.5281/zenodo.11394477} to compute \av\ and \rv\ from the tabulated reddening parameters $\xi$ (a parameter reflecting \rv) and $E$ (a parameter proportional to \av) for each of the target stars according to the ensemble of reddening curves constructed in that work. The two nearest stars (Targets 34 and 31) have quality flags $>$ 8, indicating unreliable solutions, so we adopt \av = 0.0 for these targets with no corresponding uncertainties. Our own independent analysis of the broadband SED in conjunction with $Gaia$ stellar parameters supports minimal or no extinction. Similarly, we do not adopt any \rv\ for the Near Group targets as the \av$\approx$0 implies an undefined \rv. Table \ref{tbl: reddenings} records the adopted reddening parameters and uncertainties. Column 1 is the Target number, Columns 2 and 3 are the \av\ and corresponding uncertainties, Columns 4 and 5 are the \rv\ and uncertainties, and Columns 6 and 7 are $E(B-V)$ and uncertainties (as computed from the previous columns). \av\ ranges from near zero for the nearest two targets to 1.80 mag for the most distant ones. \rv\ covers a narrow range from 2.87 to 3.86. The uncertainties are calculated straightforwardly from the uncertainties on $\xi$ and $E$. The table footnotes indicate targets with quality flags either $>8$ or $=8$ (footnotes a and b, respectively). We do not reject these targets, as they may still be valuable as part of the larger dataset, but there may be some instances in which these ``poor quality" flags explain larger deviations from the mean.

\begin{table}
    \caption{Reddening Parameters \label{tbl: reddenings}}
    \begin{center}
    \input{table3}

    \tablenotetext{a}{Targets with quality flags $>8$ in \citet{zhang2025}.}
    \tablenotetext{b}{Targets with quality flags $=8$ in \citet{zhang2025}.}
    \end{center}
\end{table}

\subsection{$E(B-V)$} \label{subsec: ebv}

\begin{figure}[h]
    \centering
    \plotone{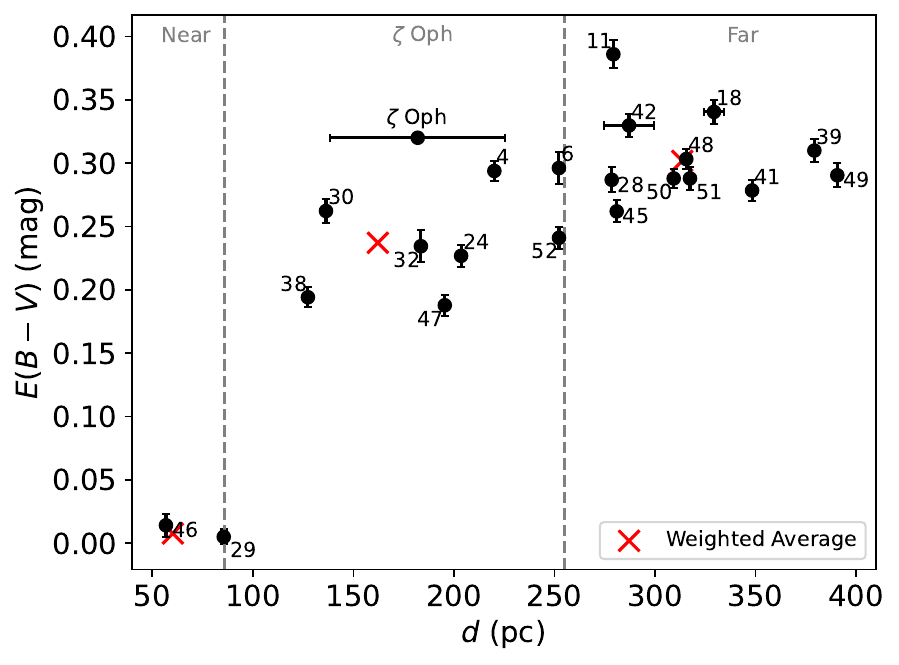}
    \caption{The target stars' $E(B-V)$ versus distance. The error bars represent the 1$\sigma$ uncertainties. The dashed lines and gray labels demarcate the three distance groups. The red $\times$'s indicate the weighted averages of each group.}
    \label{fig: ebv_dist}
\end{figure}

Figure \ref{fig: ebv_dist} plots $E(B-V)$ versus $d$ for the same targets as in Figure \ref{fig: qu-plot} (with the exception of Targets 34 and 31 which lack $E(B-V)$). The error bars represent the $1\sigma$ uncertainties as presented in Table 3. Dashed vertical lines and corresponding gray labels delineate the boundaries of the Near, $\zeta$ Oph, and Far Groups. We find that the $E(B-V)$ of targets in the Near Group cluster at zero, as with their polarization, and the weighted average is $\overline{E(B-V)} = 0.01 \pm 0.01$ mag. This indicates the near absence of dust within the Near Group, which is consistent with the minimal polarization. There is then a sudden increase in reddening between 86 pc and 127 pc to the $\zeta$ Oph Group, which has a weighted average of $\overline{E(B-V)} = 0.24 \pm 0.01$ mag. This increase in reddening from the Near to the $\zeta$ Oph Group is also consistent with the sharp increase in polarization (noted in Section \ref{sec: pols}), further indicating this as the distance range at which significant dust arises. The behavior of $E(B-V)$ with $d$ differs from the behavior in the $q$--$u$ plane in that there is no drastic change in reddening between the $\zeta$ Oph Group and Far Group ($\overline{E(B-V)} = 0.30 \pm 0.01$ mag). Instead, there is, at most, a small increase in reddening beyond 250 pc. Within the $\zeta$ Oph and Far Groups, the data appear mostly constant. This implies there is minimal dust present within these distance groups.

$\zeta$ Oph is conspicuous in its location within Figure \ref{fig: ebv_dist}. We plot $\zeta$ Oph at $E(B-V) = 0.32$ mag, as it is historically cited \citep[ex.][]{morton1975, ccm89, diplas1994}\footnote{$\zeta$ Oph is not included in the \citet{zhang2025} catalog}. This places $\zeta$ Oph at 0.08 mag greater reddening than the average of the surrounding targets. The cause of this discrepancy is unclear. If it were at $d\approx$250--300 pc, it would be of similar reddening to the targets within the Far Group, but it would then be inconsistent in its polarization, which places it firmly within the $\zeta$ Oph Group (Figure \ref{fig: qu-plot}). Systematic differences between the methods of \citet{zhang2025} and historical measurements of reddening toward O stars are possible explanations.

\subsection{\rv} \label{subsec: rv}

\begin{figure}[h]
    \centering
    \plotone{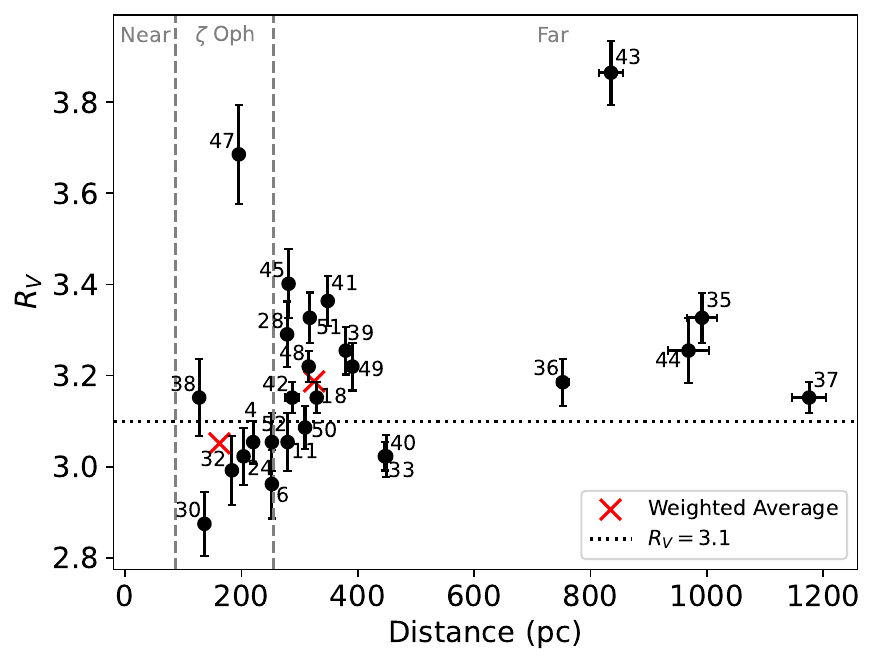}
    \caption{The target stars' \rv\ versus $d$. The red $\times$s indicate the location of the weighted averages of each distance group.}
    \label{fig: rv_dist}
\end{figure}

Figure \ref{fig: rv_dist} presents \rv\ versus $d$, including all target stars in the entire distance range as well as those from PK22 included in Figure \ref{fig: qu-plot}. The black data points represent the \rv\ for each target star, and the error bars represent the 1$\sigma$ uncertainties. All Near Group targets are omitted from the figure on account of their very low reddening. The red $\times$'s represent the weighted average \rv\ values for each distance group and their corresponding uncertainties. The gray dashed lines and labels indicate the boundaries of each distance group.

Overall, the targets have a mean \rv\ of close to 3.1, consistent with the Milky Way average, but targets in both distance groups display dispersion beyond the uncertainties. The groups have weighted means of $\overline{R_{\rm V}}$ = $3.05 \pm 0.02$ and $\overline{R_{\rm V}}$ = $3.19 \pm 0.01$ for the $\zeta$ Oph and Far Groups, respectively, indicating a slight increase in $\overline{R_{\rm V}}$. The significant dispersion in Figure \ref{fig: rv_dist} makes it difficult to draw conclusions about the behavior of \rv\ with Heliocentric distance, but the targets near $\zeta$ Oph do not display particularly anomalous \rv\ compared to those in the rest of the field. We do find that as distance increases, the dispersion appears to decrease, going from $RMS = 0.23$ in the $\zeta$ Oph group down to $RMS = 0.19$ in the Far Group. This could be explained by an increase in distance (as a proxy for increasing dust column density) leading to an increase in the number of individual dust components being sampled, and thus to a greater superposition of \rv\ and a convergence to the canonical galactic average. Particular outliers include Target 47 ($\zeta$ Oph Group) and Target 43 (Far Group), both of which are noted as having quality flags $\geq 8$ in the \citet{zhang2025} catalog. Binarity or stellar variability may explain these two outliers.

\section{Discussion} \label{sec: discussion}

\subsection{Constraining the Line of Sight Morphology} \label{subsec: dists}

\begin{figure}[h]
    \centering
    \plotone{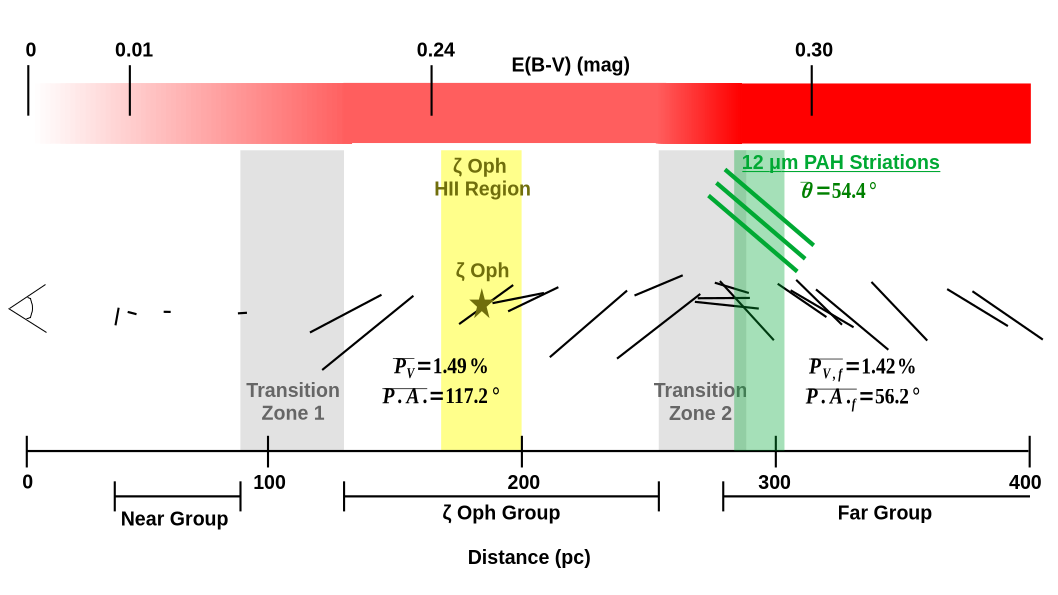}
    \caption{Schematic representation of dust and magnetic structure along the $\zeta$ Oph line of sight.}
    \label{fig: illustration}
\end{figure}

By combining the reddening and polarimetry data, we are able to produce a more complete picture of how the dust and magnetic field structure change along this line of sight. Figure \ref{fig: illustration} is a one-dimensional schematic representation of the polarization and reddening features along the sight line. The lower axis represents Heliocentric distance (out to 400 pc) with the three target distance groups marked below. Various annotations are separated vertically for clarity. The red gradient strip at the top represents the change in $E(B-V)$ with distance, and the 23 black line segments depict the \pv\ and $P.A.$ of the individual targets, as in Figure \ref{fig: skypol-subfore}. The two gray vertical bands labeled ``Transition Zone 1" and ``Transition Zone 2" represent the regions between distance groups where changes in polarization and reddening occur. 

\subsubsection{Reddening Components} \label{subsubsec: red dists}

The weighted average reddenings of the three distance groups (Near, $\zeta$ Oph, and Far up to 400 pc) are $\overline{E(B-V)} =$ 0.01 mag, 0.24 mag, and 0.30 mag, respectively. These are labeled along the top red gradient of Figure \ref{fig: illustration}. The data in Figure \ref{fig: ebv_dist} indicate that there is a significant jump in the reddening, and therefore the dust column density, between the Near and $\zeta$ Oph Groups (within Transition Zone 1). This likely marks the first significant instance of dust present along this sight line. There is another jump between the $\zeta$ Oph and Far Groups (Transition Zone 2), though it is significantly smaller than that of the first. The smaller average increase in reddening within Transition Zone 2 (as opposed to Transition Zone 1) indicates a smaller dust column density within this interval. Within the $\zeta$ Oph and Far Groups, the reddening remains mostly constant, with some scatter, suggesting there may be little to no dust between and background to the Transition Zones. This behavior is represented by the two darkening red gradients within the Near Group and Transition Zones 1 and 2 followed by two constant red shades in the $\zeta$ Oph and Far Groups in Figure \ref{fig: illustration}. We find no compelling evidence for coherent, across-field variations in $E(B-V)$, as there appear to be mostly small random variations with angular distance. Further evidence of this comes from the reddenings of Targets 28 and 29, both of which are displaced from the rest of the field by $\Delta \theta > 0.5$\degr\ but are consistent with the reddenings of other targets within their distance groups. While there are variations in reddening within distance groups, it is appropriate to characterize the extinction as primarily a function of distance and approximately uniform across the $\sim$1\degr\ field. Based on this behavior, we characterize the dust distribution along the sight line as existing primarily within the two Transition Zones and minimal or absent between the two and background to $\sim 300$ pc.

The scatter in \rv\ prevents any similar widespread inferences about dust grain sizes within the Transition Zones. Instead, as discussed in Section \ref{subsec: rv}, \rv\ varies significantly---and apparently randomly---across these sight lines. This is consistent with a number of previous studies \citep[see][]{johnsonmorgan1955,fitzpatrick1988,ccm89} which have concluded that real variations in \rv\ occur on small angular scales. An example of this is a pair of targets in the \citet{ccm89} study \citep[data from][]{fitzpatrick1988}: HD 37903 and HD 38087 with $\Delta \theta = 0.35$\degr, $E(B-V) = 0.35$ and 0.33 mag, and \rv\ $ = 4.11$ and 5.30, respectively. The similarity in $E(B-V)$ indicates that these are probing similar columns of dust, but the targets have notably different \rv. Similarly, Targets 51 and 52 have $\Delta \theta = 0.40$\degr, $E(B-V) =$ 0.29 and 0.24 mag, and \rv\ = 3.33 and 3.05. Real variations in \rv\ occur within the Milky Way, and the galactic average is representation of many different dust populations and environments \citep{ccm89, f19, g23}. Along sight lines with very low $E(B-V)$, these variations could be enhanced, as the stochastic effects of sampling individual dust size/composition populations may dominate the extinction and polarization characteristics. Over small volumes of the ISM, it is possible that only a limited range of the canonical power-law dust grain size distribution is being sampled. To our knowledge this possibility has not been directly proposed or investigated.

\subsubsection{Magnetic Structures} \label{subsubsec: mag dists}

Figure \ref{fig: illustration} presents the polarization vectors of all of the targets present in Figure \ref{fig: qu-plot} (those with $d < 400$ pc) in one dimension to better examine their relationships to other features with distance. Section \ref{sec: pols} described the distinct changes in the strength and orientation of the $B$ field along this sight line. The first such instance is the previously noted increase in $q$ and $u$  between Targets 29 and 38 ($d = 86$ pc and 127 pc) within Figure \ref{fig: qu-plot}. This indicates that the first polarizing dust component (i.e., the first group of predominantly aligned grains), and therefore the first of the measured magnetic structures, arises within that short 41 pc range denoted by Transition Zone 1. The $\zeta$ Oph group, with an average $\overline{P.A.} = 117.2$\degr\ and $\overline{P_{\rm V}} = 1.49$\%, all reflect the polarization of this magnetic structure contained within Transition Zone 1. Over the range of the $\zeta$ Oph Group, the targets' polarization percentages and angles vary only a small amount ($RMS_{P}$ = 0.41\% and $RMS_{P.A.}$ = 8.62\degr), indicating either poor grain alignment or an ambient \textbf{B} field parallel to the line of sight, if dust is present. Section \ref{subsubsec: red dists} suggests the most likely scenario is that there is minimal to no dust within the $\zeta$ Oph Group to cause additional polarization. The aligned dust grains foreground to 252 pc must therefore be constrained to Transition Zone 1.

While $\zeta$ Oph lies within the middle of this distance group, the group itself spans a distance range of $d = $ 127--252 pc ($\Delta d = 125$ pc), far larger than the approximately 16 pc radius of $\zeta$ Oph's \ion{H}{2} region \citep[approximately 5\degr\ radius in the H$\alpha$ map from][represented as the yellow vertical band in Figure \ref{fig: illustration}]{finkbeiner2003}. The location of the polarizing grains is also at least 39 pc foreground to the \ion{H}{2} region. This indicates that neither the alignment of the dust grains within Transition Zone 1 nor the lack of alignment within the $\zeta$ Oph Group are predominantly due to $\zeta$ Oph's radiative influence, but rather to the structure of the ambient \textbf{B} field. One potential cause of this sudden increase in dust alignment within Transition Zone 1 is the presence of the Local Bubble (LB) wall. The LB is characterized by a cavity of low-density ISM expanding into higher-density, cooler gas and dust. The boundary, or wall, of the LB marks a sharp increase in both reddening and polarization with distance along a line of sight \citep[see][]{leroy1999, lallement2003}. While the shape of the LB is irregular, the distance to the wall in the direction of $\zeta$ Oph is likely $\lesssim 100$ pc \citep{lallement2003, medan2019,pelgrims2020, oneill2024}. \citet{leroy1999} and \citet{gontcharov2019} found that the wall can be defined as the distance at which \pv = 0.1\%. The Near Group ranges from \pv =  $0.14 \pm 0.03$--$0.29 \pm 0.10$\%, indicating that the wall may lie at or foreground to the 36--86 pc distance range of the group. This implies that it is either not the wall of the LB that we are measuring with the sudden increase in polarization in Transition Zone 1, or that the polarization of the wall is characterized differently in this direction.

The transition to the Far Group between 252 pc and 287 pc ($\Delta d = 35$ pc, Transition Zone 2), and the corresponding change in polarization to $\overline{P_{{\rm V},f}} = 1.42\%$ and $\overline{P.A_{f}} = 56.2$\degr\ (after foreground-subtraction of the $\zeta$ Oph Group), represents a systematic, large-scale rotation of the projected orientation of the ambient magnetic field. This is easily seen in Figure \ref{fig: illustration}, as the individual polarization line segments rotate by $\Delta \overline{P.A.} = 61$\degr. We suggest that this second population of aligned grains is the ``mid-distance dust component" measured by PK22 and suspected to be between 200--300 pc. The average Far Group polarizations presented are limited to the targets contained between 287--391 pc (while the Far Group extends beyond 400 pc, there is increasing scatter within the $q$--$u$ plane at larger distances, and the reddening data indicate a lack of dust at these distances). Although there is not a substantial increase in \pv\ within Transition Zone 2, this does not necessarily imply either a lack of dust or a lack of aligned grains. Rather, there is a significant rotation in the $q$--$u$ plane which represents a substantial vector difference between the two groups of targets in Figure \ref{fig: qu-plot}. Furthermore, the alignment of the foreground-subtracted Far Group $\overline{P.A.}$ with the average orientation, $\overline{\theta} = 54.4$\degr\ (PK22), of the 12 \micron\ PAH striations (represented in green) suggests that the PAH-emitting grains are spatially coincident with those responsible for the polarization. The three target stars within Transition Zone 2 do not share the orientation of the striations, indicating that the striations arise background to 281 pc (Target 45). Placing the PAH striations beyond $\sim$300 pc seems unlikely due to the lack of dust (reddening) beyond this point. Therefore, our polarization and reddening data strongly suggest---but do not conclusively establish---that the PAH-emitting and polarizing grains are spatially coincident within this $\sim$20 pc range and may, in fact, be the same population.

\subsection{Polarization Efficiency}

\begin{figure}
    \centering
    \plotone{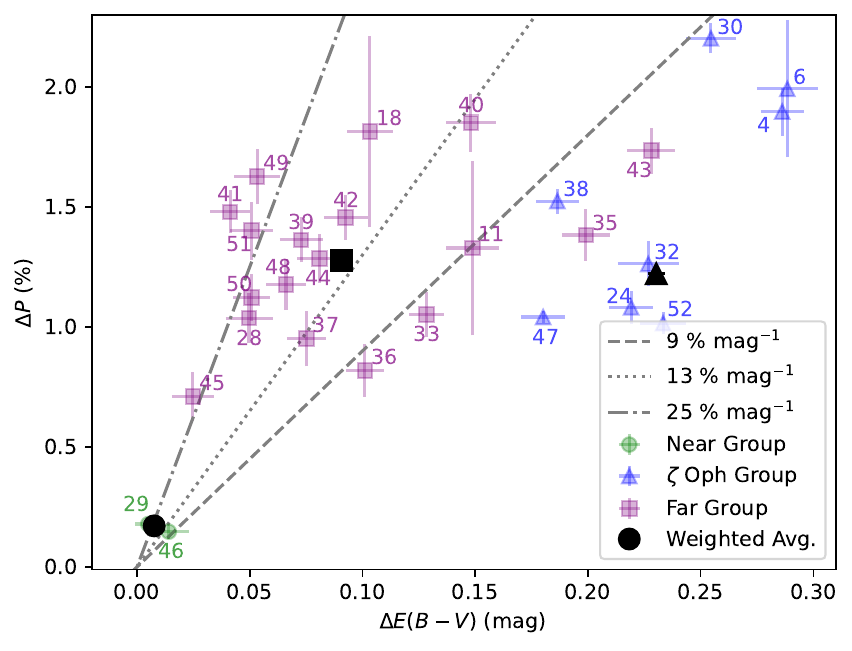}
    \caption{Target stars' polarization efficiencies. Colored points (green, blue, and purple for Near, $\zeta$ Oph, and Far Groups, respectively) represent individual targets while black points represent the weighted averages of each distance group. The dashed line is the polarization efficiency limit determined by \citet{Serkowski1975}, and the dotted line is the limit from \citet{panopoulou2019}. The dash-dotted line represents 25\% mag$^{-1}$ as a reference.}
    \label{fig: p_ebv}
\end{figure}

Figure \ref{fig: p_ebv} presents $P_{{\rm V},f}$ (or \pv\ for the Near and $\zeta$ Oph Groups) versus $\Delta E(B-V)$ (foreground-subtracted reddening) as a way to examine polarization efficiency for each target star. The targets included in the figure are the same as those represented in Figure \ref{fig: ebv_dist}. The reddening data for the $\zeta$ Oph and Far Group targets are foreground subtracted by the weighted averages of the Near and $\zeta$ Oph Groups, respectively. The polarization data for only the Far Group are similarly foreground subtracted, as the polarization of the Near Group (i.e., the foreground of the $\zeta$ Oph Group) scatters near zero. The symbols correspond to the three distance groups, as in previous figures, and the green, blue, and purple colors also differentiate the groups for ease of viewing. The bold symbols represent the weighted averages of each group. The error bars indicate the $1 \sigma$ uncertainties. The dashed and dotted lines represent the \citet{Serkowski1975} (\pebv = 9\% mag$^{-1}$) and \citet{panopoulou2019} (\pebv = 13\% mag$^{-1}$) maximum polarization efficiency relationships, respectively.

The polarization efficiencies of the $\zeta$ Oph Group show a quasi-linear increase with reddening. This is expected, as an increase in reddening, and therefore a higher column density of dust, would correspond to an increase in polarization if the magnetic field $P.A.$ remains consistent over this distance. The Far Group shows a more scattered distribution in polarization efficiency. Most of the targets lie above the traditional Serkowski limit, many above the 9--13\% mag$^{-1}$ limits, with a weighted average of $14.1 \pm 0.5$\% mag$^{-1}$. This provides further evidence that super-Serkowski efficiencies are not uncommon in the diffuse ISM. Both \citet{planck2015} and \citet{panopoulou2019} came to similar conclusions, setting and confirming that the upper limit of polarization efficiency could be \pebv $\geq$ 13\% mag$^{-1}$. \citet{panopoulou2019} also concluded that their polarization efficiencies were highly dependent on accurate reddening determinations.

A physical explanation for these high polarization efficiencies could be that the dust grains responsible have a different size or composition than the ISM average. \citet{panopoulou2019} investigated a region with anomalous infrared polarized emission in the \textit{Planck} survey and found that it also displayed a high optical polarization efficiency. They found that the wavelength dependence of optical polarization was typical for the ISM and concluded that the sight line did not contain an anomalous dust composition or size distribution. Because we do not have multi-wavelength polarimetry, we instead examine \rv\ as a method of investigating whether grain sizes  along $\zeta$ Oph sight lines are anomalous. The well-established, positive relation between \rv\ and the wavelength of maximum polarization \citep[$\lambda_{\rm max}$;][]{Serkowski1975} indicates that larger grains are usually more aligned than smaller grains, leading to a typical $\lambda_{\rm max}$ in the optical. If the size distribution of aligned dust shifts towards smaller grains, $\lambda_{\rm max}$ and \rv\ decrease as well, and the enhanced polarizing ability of smaller grains could lead to elevated \pebv\ \citep{clayton1988, larson2000}. In our data we find no correlation between \rv\ and \pebv, leading us to reject this as the cause of the high polarization efficiencies within the $\zeta$ Oph sight lines. Thus, the high \pebv\ is likely not a result of abnormal grain sizes or compositions but it, like \rv, has measurable variations on angular scales of less than a few arcminutes.

It is also possible that the dust alignment, and high \pebv, measured by the Far Group (the dust contained within Transition Zone 2) is caused by the rotation of the projected \textbf{B} field to an orientation which lies primarily in the plane of the sky. \citet{skalidis2024} concluded there was a correlation between high polarization efficiency and low reddening, with \pebv\ being suppressed beyond $\sim$0.5 mag due to line-of-sight depolarization effects. While none of our targets have $E(B-V) > 0.50$ mag, even prior to foreground subtraction, we do, in fact, see line-of-sight depolarization between the $\zeta$ Oph and Far Groups over the small reddening range of only $\Delta E(B-V) \simeq 0.3$ mag. The true maximum polarization efficiency is often masked by overlapping dust and magnetic structures. Measuring the true maximum polarization efficiency therefore requires a proper removal of foreground components. This indicates that in order to discover areas of high polarization efficiency, studies must be able to probe limited volumes of the ISM where $E(B-V)$ is small to mitigate line-of-sight effects. This conclusion is also consistent with PK22's suggestion that the 12 \micron\ PAH striations and corresponding super-Serkowski \pebv\ are contained within a very small fractional ISM volume. 

PK22 also speculated that $\zeta$ Oph's UV radiation was either sputtering or spinning up the grains responsible for the high \pebv\ while simultaneously illuminating the the PAH striations. There have been multiple distance estimates for $\zeta$ Oph, placing it in the range of 112 pc \citep{vanleeuwen2007} to 230 pc \citep{siebenmorgen2020}, with the $182^{+53}_{-34}$ pc we adopt from \citet{Bailer-Jones2018} in between. Even adopting the most extreme (230 pc) distance would place $\zeta$ Oph's \ion{H}{2} region, at its farthest extent, at 250 pc. \citet{draine1986} established that beyond the extent of $\zeta$ Oph's \ion{H}{2} region, the contribution it makes to the UV field is quickly comparable to or smaller than the influence of the interstellar radiation field (ISRF). By building on the PK22 observations, we have been able to further constrain the location of the PAH striations to a distance of $d \simeq$ 281--300 pc, which places them significantly in the background of $\zeta$ Oph's radiative influence. This indicates they are being illuminated by some other source, not necessarily one with a UV intense radiation field like $\zeta$ Oph. The general ISRF is sufficient to excite the 7.7 \micron\ and 11.3 \micron/12.7 \micron\ PAH features (in the Spitzer 8 \micron\ and WISE 12 \micron\ bands), even at low intensities \citep{draine2007, dale2023}. The $I_{8 \micron}$/$I_{4.5 \micron}$ ($\approx 0.20$) and $I_{8 \micron}$/$I_{24 \micron}$ ($\approx 20$) specific intensity ratios of the striations calculated by PK22 could be consistent with a range of starlight intensities and PAH mass fractions, but in all combinations it is possible the ISRF is responsible for the PAH emission from the striations. As there is little to no dust within the radiative influence of $\zeta$ Oph, our data do not directly contradict PK22's suggestion that stars of similar radiative power could be responsible for high \pebv, but simply dismisses this as the cause within this particular sight line.

\section{Conclusions} \label{sec: conclusions}

Building on the work of \citet{picconekobulnicky2022}, we conducted a $V$ band polarimentric and photometric study of 25 target stars within a 50\arcmin\ radius of $\zeta$ Oph and with \textit{Gaia} DR3 distances of $d = $ 36--1176 pc. Utilizing these precise distances, we have mapped the characteristics and locations of the dust and magnetic field components unaffected by line-of-sight superposition. There are at least two distinct dust populations and projected \textbf{B} field orientations: Transition Zone 1 ($d = $ 86--127 pc) and Transition Zone 2 ($d = $ 252--287 pc). Transition Zone 1 corresponds to a significant increase in \pv\ ($\overline{P.A.} = 117.2$\degr\ and $\overline{P_{\rm V}} = 1.49$\%) and $E(B-V)$ (0.24 mag) between the Near and $\zeta$ Oph Groups, and Transition Zone 2 represents a rotation in $P.A.$ ( $\overline{P.A_{f}} = 56.2$\degr\ and $\overline{P_{{\rm V},f}} = 1.42\%$) and a slight increase of $\Delta E(B-V) = $ 0.06 mag between the $\zeta$ Oph and Far Groups. Within these distance groups (foreground, between, and background to the two Transition Zones), polarization and reddening are mostly constant, indicating minimal dust.

The average polarization efficiency of the Far Group (impacted by the dust within Transition Zone 2) is \pebv\ = 14.1\%, with most targets displaying even higher values. These are consistent with suggestions that super-Serkowski polarization efficiencies may be common in the ISM, but observing them requires precise foreground subtraction to mitigate the suppression of \pebv\ by line-of-sight depolarization. Because \rv\ appears mostly consistent with the galactic average within the Far Group (in contrast to PK22's findings), we suggest that the high \pebv\ is a result of favorable magnetic field geometry and/or enhanced alignment rather than unusual dust size or composition. Alignment of the 12 \micron\ PAH striations (seen in the WISE image) with the foreground-subtracted $P.A.$ of the Far Group indicates the dust responsible for the high polarization efficiency and dust producing the PAH emission may be spatially coincident within $d \simeq$ 281--300 pc. While PK22 speculated that UV radiation from $\zeta$ Oph may be responsible for the high \pebv\ and simultaneously the illumination of the PAH striations, our improved distance range for this dust population places it outside even the most extreme estimate of $\zeta$ Oph's radiative influence. Despite this, our data are consistent with the conclusions of PK22 that super-Serkowski polarization efficiencies may originate in small fractional volumes of the ISM, given that we can constrain the dust responsible to a roughly 50 pc range. For low-reddening, diffuse sight lines, real variations in \rv\ and \pebv\ may be a result of stochastic sampling of small ISM volumes where the dust grain size and/or composition distributions are not well mixed.

Without accurate \textit{Gaia} distances and foreground subtraction, we would have been unable to distinguish many of these important characteristics of the dust populations and \textbf{B} field components from one another. Further studies of polarization and reddening require datasets well-sampled in distance to properly account for changes on small scales and line-of-sight depolarization.

\begin{acknowledgments}
This research has made use of the NASA/IPAC Infrared Science Archive, which is funded by the National Aeronautics and Space Administration and operated by the California Institute of Technology. NASA grants 80NSSC22K0484 and 80NSSC21K1847 supported this work. We thank the anonymous referee for helpful suggestions for improving this work, Ashley Piccone and Evan M. Cook for assistance in obtaining the polarimetric data at WIRO, and Ethan Cotter for assistance with writing the code for our independent analysis of extinction.
\end{acknowledgments}

\facilities{WIRO, RBO, \textit{Gaia}, WISE}

\software{IRAF \citep{tody1986}}

\bibliography{bib}{}
\bibliographystyle{aasjournal}

\end{document}

%% file: table1.tex
\begin{tabular}{c c c c c c c c c c}
\hline\hline 
Target & Group & {\it Gaia} DR3 ID & R.A. (2016) & Decl. (2016) & Distance & T$_{\text{eff}}$ & log g & [M/H] & RUWE  \\ 
  &   &   & (deg) & (deg) & (pc)  & (K) &   &   &    \\ 
\hline 
28 & Far & 4332097670528328064 & 248.4634 & -10.7561 & 278$\pm$1.3 & 4648 & 1.97 & -0.22 & 1.00 \\ 
29 & Near & 4332104228941248640 & 248.5296 & -10.6779 & 86$\pm$0.1 & 5046 & 3.30 & -0.42 & 0.93 \\ 
30 & $\zeta$ Oph & 4338134023726756224 & 249.1830 & -10.2907 & 136$\pm$0.4 & 5328 & 4.82 & 0.15 & 1.38 \\ 
31 & Near & 4337323133900854400 & 249.1829 & -10.9741 & 42$\pm$0.04 & 3230 & 3.91 & -0.37 & 1.10 \\ 
32 & $\zeta$ Oph & 4337346842120155392 & 249.2510 & -10.6659 & 184$\pm$0.6 & 5587 & 4.03 & -0.23 & 1.15 \\ 
33 & Far & 4337345570809807360 & 249.2692 & -10.7342 & 446$\pm$6.3 & 5432 & 3.93 & 0.19 & 1.81 \\ 
34 & Near & 4337345708248760448 & 249.2930 & -10.7187 & 36$\pm$0.03 & 3554 & 4.52 & -0.44 & 1.14 \\ 
35 & Far & 4338130209794120320 & 249.3393 & -10.3717 & 992$\pm$26.2 & 3821 & 0.34 & -0.66 & 1.18 \\ 
36 & Far & 4338141411068864384 & 249.4136 & -10.2063 & 752$\pm$9.8 & 4136 & 0.45 & -1.02 & 1.17 \\ 
37 & Far & 4337329146854784640 & 249.4326 & -10.8356 & 1176$\pm$29.1 & 4484 & 1.70 & -0.21 & 1.32 \\ 
38 & $\zeta$ Oph & 4337375154544565504 & 249.5429 & -10.4570 & 127$\pm$0.4 & 5151 & 4.50 & 0.10 & 1.94 \\ 
39 & Far & 4338139486923481728 & 249.5683 & -10.1910 & 379$\pm$2.3 & 4618 & 1.98 & 0.00 & 1.00 \\ 
40 & Far & 4337361097113403776 & 249.6516 & -10.5616 & 449$\pm$2.6 & 6286 & 4.21 & 0.02 & 0.97 \\ 
41 & Far & 4337281970933936384 & 249.6561 & -10.8482 & 348$\pm$2.0 & 4665 & 1.85 & -0.37 & 0.97 \\ 
42 & Far & 4337354469981944192 & 249.6606 & -10.7105 & 287$\pm$12.4 & 6350\tablenotemark{a} & 3.73\tablenotemark{a} & -0.06\tablenotemark{a} & 7.88 \\ 
43 & Far & 4337275064626457472 & 249.6644 & -11.0352 & 835$\pm$21.2 & 3831 & 0.72 & -0.35 & 1.33 \\ 
44 & Far & 4337368213877372672 & 249.6655 & -10.4824 & 968$\pm$35.0 & 4353 & 1.67 & -0.14 & 2.32 \\ 
45 & Far & 4337368832352669568 & 249.7037 & -10.4390 & 281$\pm$1.8 & 4503 & 1.90 & 0.07 & 1.22 \\ 
46 & Near & 4337308702810407040 & 249.7296 & -10.7331 & 57$\pm$0.06 & 4105 & 4.89 & 0.06 & 1.15 \\ 
47 & $\zeta$ Oph & 4337478130679012224 & 249.7460 & -10.2646 & 195$\pm$0.9 & 4631 & 1.98 & -0.18 & 1.02 \\ 
48 & Far & 4337265306461090944 & 249.7506 & -11.0029 & 316$\pm$2.1 & 6167 & 4.05 & -0.48 & 1.07 \\ 
49 & Far & 4337477954582912640 & 249.7563 & -10.2676 & 391$\pm$2.4 & 6028 & 4.06 & -0.15 & 1.05 \\ 
50 & Far & 4337367247506569856 & 249.7594 & -10.5196 & 309$\pm$2.0 & 7302\tablenotemark{a} & 4.04\tablenotemark{a} & -0.81\tablenotemark{a} & 1.22 \\ 
51 & Far & 4337477885863436288 & 249.7782 & -10.2804 & 317$\pm$1.5 & 5778 & 4.31 & 0.27 & 1.00 \\ 
52 & $\zeta$ Oph & 4337314367869226624 & 249.8716 & -10.6730 & 252$\pm$1.4 & 5783 & 3.81 & -0.28 & 1.25 \\ 
\end{tabular}

%% file: table2.tex
\begin{tabular}{c c c c c c c} 
\hline\hline 
Target &  $q$ & $u$ & $P_{\rm V}$ & $P.A.$ & $P_{{\rm V},f}$ & $P.A._f$ \\ 
  & (\%) & (\%) & (\%) & (deg.) & (\%) & (deg.) \\ 
\hline 

34 & 0.28$\pm$0.10 & -0.12 $\pm$0.10 & 0.28$\pm$0.10 & 167.90$\pm$9.92 & \nodata & \nodata \\ 
31 & -0.21$\pm$0.16 & 0.10 $\pm$0.16 & 0.17$\pm$0.16 & 77.60 $\pm$44.09 & \nodata & \nodata \\ 
46 & -0.15$\pm$0.03 & 0.00 $\pm$0.03 & 0.14$\pm$0.03 & 89.40 $\pm$6.80 & \nodata & \nodata \\ 
29 & -0.18$\pm$0.02 & -0.02$\pm$0.02 & 0.18$\pm$0.02 & 93.10 $\pm$3.02 & \nodata & \nodata \\ 
\hline
38 & -1.03$\pm$0.05 & -1.13$\pm$0.05 & 1.52$\pm$0.05 & 113.80$\pm$1.01 & \nodata & \nodata \\ 
30 & -0.81$\pm$0.06 & -2.05$\pm$0.06 & 2.20$\pm$0.06 & 124.20$\pm$0.83 & \nodata & \nodata \\ 
32 & -0.59$\pm$0.09 & -1.12$\pm$0.09 & 1.26$\pm$0.09 & 121.10$\pm$2.19 & \nodata & \nodata \\
47 & -0.99$\pm$0.02 & -0.34$\pm$0.02 & 1.04$\pm$0.02 & 99.40 $\pm$0.67 & \nodata & \nodata \\ 
52 & -0.79$\pm$0.05 & -0.63$\pm$0.05 & 1.01$\pm$0.05 & 109.30$\pm$1.27 & \nodata & \nodata \\ 
\hline
28 & -1.80$\pm$0.02 & -1.23$\pm$0.02 & 2.18$\pm$0.02 & 107.20$\pm$0.35 & 1.03$\pm$0.10 & 90.25$\pm$2.36 \\ 
45 & -1.39$\pm$0.03 & -0.89$\pm$0.03 & 1.65$\pm$0.03 & 106.30$\pm$0.49 & 0.70$\pm$0.11 & 75.87$\pm$3.68 \\ 
42 & -0.89$\pm$0.05 & 0.23 $\pm$0.05 & 0.92$\pm$0.05 & 82.80 $\pm$1.45 & 1.45$\pm$0.11 & 47.52$\pm$2.18 \\
50 & -1.33$\pm$0.04 & -0.25$\pm$0.04 & 1.35$\pm$0.04 & 95.40 $\pm$0.88 & 1.12$\pm$0.11 & 60.25$\pm$2.68 \\
48 & -0.99$\pm$0.07 & -0.07$\pm$0.07 & 0.99$\pm$0.07 & 92.00 $\pm$1.88 & 1.17$\pm$0.12 & 50.71$\pm$2.93 \\
51 & -1.62$\pm$0.08 & -0.11$\pm$0.08 & 1.62$\pm$0.08 & 92.00 $\pm$1.40 & 1.40$\pm$0.13 & 63.85$\pm$2.53 \\
41 & -0.95$\pm$0.04 & 0.24 $\pm$0.04 & 0.98$\pm$0.04 & 82.80 $\pm$1.01 & 1.48$\pm$0.11 & 48.73$\pm$2.07 \\
39 & -1.56$\pm$0.03 & -0.12$\pm$0.03 & 1.57$\pm$0.03 & 92.20 $\pm$0.48 & 1.36$\pm$0.11 & 63.02$\pm$2.08 \\ 
49 & -1.58$\pm$0.08 & 0.18 $\pm$0.08 & 1.59$\pm$0.08 & 86.70 $\pm$1.40 & 1.62$\pm$0.13 & 60.15$\pm$2.17 \\ 
33 & -1.22$\pm$0.04 & -0.28$\pm$0.04 & 1.25$\pm$0.04 & 96.40 $\pm$0.92 & 1.05$\pm$0.11 & 58.06$\pm$2.85 \\
40 & -1.06$\pm$0.09 & 0.60 $\pm$0.09 & 1.22$\pm$0.09 & 75.10 $\pm$2.00 & 1.85$\pm$0.13 & 49.60$\pm$2.07 \\ 
36 & -1.42$\pm$0.06 & -0.74$\pm$0.06 & 1.60$\pm$0.06 & 103.70$\pm$1.01 & 0.81$\pm$0.12 & 71.80$\pm$3.68 \\
43 & -1.22$\pm$0.05 & 0.45 $\pm$0.05 & 1.30$\pm$0.05 & 79.90 $\pm$1.01 & 1.73$\pm$0.11 & 52.72$\pm$1.82 \\
44 & -1.40$\pm$0.06 & -0.11$\pm$0.06 & 1.41$\pm$0.06 & 92.20 $\pm$1.08 & 1.28$\pm$0.12 & 60.00$\pm$2.49 \\
35 & -1.98$\pm$0.05 & -0.56$\pm$0.05 & 2.05$\pm$0.05 & 97.90 $\pm$0.60 & 1.38$\pm$0.11 & 75.68$\pm$2.03 \\ 
37 & -0.78$\pm$0.08 & -0.27$\pm$0.08 & 0.82$\pm$0.08 & 99.60 $\pm$2.76 & 0.94$\pm$0.13 & 45.51$\pm$3.88 \\ 

\end{tabular}

%% file: table3.tex
\begin{tabular}{c c c c c c c} 
\hline\hline 
Target & $A_{\rm V}$ & $A_{\rm V}$ unc. & $R_{\rm V}$ & $R_{\rm V}$ unc. & $E(B-V)$ & $E(B-V)$ unc. \\ 
 & (mag) & (mag) &  & & (mag) & (mag)  \\ 
\hline 

34\tablenotemark{a} & 0.00 & \nodata & \nodata & \nodata & 0.00 & \nodata \\ 
31\tablenotemark{a} & 0.00 & \nodata & \nodata & \nodata & 0.00 & \nodata \\ 
46 & 0.02 & 0.01 & \nodata & \nodata & 0.01 & 0.01 \\ 
29\tablenotemark{b} & 0.03 & 0.03 & \nodata & \nodata & 0.01 & 0.01 \\ 
\hline

38 & 0.61 & 0.02 & 3.15 & 0.08 & 0.19 & 0.01 \\ 
30 & 0.75 & 0.02 & 2.87 & 0.07 & 0.26 & 0.01 \\ 
32 & 0.70 & 0.03 & 2.99 & 0.08 & 0.23 & 0.01 \\ 
47\tablenotemark{b} & 0.69 & 0.02 & 3.69 & 0.11 & 0.19 & 0.01 \\ 
52 & 0.74 & 0.02 & 3.05 & 0.06 & 0.24 & 0.01 \\ 
\hline

28\tablenotemark{b} & 0.94 & 0.03 & 3.29 & 0.07 & 0.29 & 0.01 \\
45\tablenotemark{b} & 0.89 & 0.02 & 3.40 & 0.08 & 0.26 & 0.01 \\ 
42 & 1.04 & 0.03 & 3.15 & 0.03 & 0.33 & 0.01 \\ 
50 & 0.89 & 0.02 & 3.09 & 0.05 & 0.29 & 0.01 \\ 
48 & 0.98 & 0.02 & 3.22 & 0.03 & 0.30 & 0.01 \\ 
51 & 0.96 & 0.03 & 3.33 & 0.06 & 0.29 & 0.01 \\ 
41\tablenotemark{b} & 0.94 & 0.02 & 3.36 & 0.06 & 0.28 & 0.01 \\ 
39\tablenotemark{b} & 1.01 & 0.02 & 3.25 & 0.05 & 0.31 & 0.01 \\ 
49 & 0.94 & 0.03 & 3.21 & 0.05 & 0.29 & 0.01 \\ 
33 & 1.10 & 0.02 & 3.02 & 0.03 & 0.37 & 0.01 \\ 
40 & 1.16 & 0.03 & 3.02 & 0.05 & 0.39 & 0.01 \\ 
36 & 1.08 & 0.02 & 3.19 & 0.05 & 0.34 & 0.01 \\ 
43\tablenotemark{a} & 1.80 & 0.02 & 3.86 & 0.07 & 0.47 & 0.01 \\ 
44 & 1.04 & 0.02 & 3.25 & 0.07 & 0.32 & 0.01 \\ 
35\tablenotemark{a} & 1.45 & 0.02 & 3.33 & 0.05 & 0.44 & 0.01 \\ 
37 & 0.98 & 0.02 & 3.15 & 0.03 & 0.31 & 0.01 \\ 
\end{tabular}